\documentclass[aps,prl,preprint,superscriptaddress]{revtex4-2}
\usepackage{graphicx}
\usepackage{dcolumn}
\usepackage{bm}
\usepackage{hyperref}
\usepackage{xcolor}
\usepackage{soul}
\usepackage{amsmath}
\usepackage{amssymb}

\let\emptyset\varnothing

\usepackage[textsize=scriptsize]{todonotes}

\begin{document}

\title{Community Detection in Directed Weighted Networks using Voronoi Partitioning}



\author{Botond Molnár}
\email{botond.molnar@ubbcluj.ro}
\affiliation{Faculty of Mathematics and Computer Science, Babeș-Bolyai University, Cluj-Napoca, Romania}
\affiliation{Faculty of Physics, Babeș-Bolyai University, Cluj-Napoca, Romania}%
\affiliation{Transylvanian Institute of Neuroscience, Cluj-Napoca, Romania}
\author{Ildikó-Beáta Márton}
\affiliation{Faculty of Mathematics and Computer Science, Babeș-Bolyai University, Cluj-Napoca, Romania}
\author{Szabolcs Horv\'at}
\email{horvat@mpi-cbg.de}
\affiliation{Max Planck Institute for Cell Biology and Genetics, Dresden, Germany}
\affiliation{Center for Systems Biology Dresden, Germany}
\author{Mária Ercsey-Ravasz}%
\email{maria.ercsey@ubbcluj.ro}
\affiliation{Faculty of Physics, Babeș-Bolyai University, Cluj-Napoca, Romania}
\affiliation{Transylvanian Institute of Neuroscience, Cluj-Napoca, Romania}

\date{\today}

\begin{abstract}
Community detection is a ubiquitous problem in applied network analysis, yet efficient techniques do not yet exist for all types of network data. Most techniques have been developed for undirected graphs, and very few exist that handle directed and weighted networks effectively. Here we present such an algorithm based on Voronoi partitionings. As an added benefit, this method can directly employ edge weights that represent lengths, in contrast to algorithms that operate with connection strengths, requiring ad-hoc transformations of length data. We demonstrate the method on inter-areal brain connectivity, air transportation networks, as well as on randomly generated benchmark networks. The algorithm can handle dense graphs where weights are the main factor determining communities. The hierarchical structure of networks can also be detected, as shown for the brain. Its time efficiency is comparable with other state-of-the-art algorithms, the most costly part being Dijkstra’s shortest paths algorithm ($\mathcal{O}(|E| + |V|\log |V|)$).
\end{abstract}


\maketitle


\section{Introduction}
\label{s:introduction}
Many systems studied in social sciences, biology, neuroscience, information technology, etc.\ possess a connectivity structure and can be modelled as complex networks. Detecting communities---tightly connected groups of network nodes---is a common need in these fields. The connections in such networks may have directionality, and may also have attributes such as a \textit{strength} or \textit{length} or even both, which must be taken into account when performing community detection. These connection weights are particularly important in dense networks, where the network topology (i.e.\ the mere absence or presence of connections) may not carry as much information as the weights.

Many community detection algorithms were developed for undirected and unweighted networks \cite{fortunato, newman_pnas, louvain, newman_girvan, reichard_bornholdt, girvan_newman, infomap, VoronoiNJP}. A few of these also work with weighted undirected networks \cite{louvain, infomap, hierarchicalClustering}, but there are very few algorithms that can efficiently find communities in weighted directed networks \cite{infomap, SBM_HOLLAND1983109,Traag2019}. One of these state of the art algorithms, Infomap, is probably the best performing and most frequently applied in these types of networks. It is based on minimizing the so-called map equation (a cost function) with a stochastic method \cite{infomap}. Stochastic block modeling can also deal with directed weighted networks, but the algorithm is complex and computationally intensive, and has the disadvantage that one needs to define the desired number of clusters in advance  \cite{SBM_HOLLAND1983109}.

Community detection methods that directly maximize the modularity measure can in principle also be applied to directed networks, by using the directed generalization of modularity \cite{Leicht2008}. However, there are few practical implementations of such directed methods, with one rare example being the \texttt{leidenalg} package \cite{Traag2019,leidenalg}.

Here, we present a novel community detection method applicable to weighted directed networks based on the Voronoi partitioning of network nodes \cite{VoronoiNJP}. In this approach, a connection length is defined based on both the weight and the local topology of links, which allows for computing pairwise distances between the nodes in both directions.  The generator points of the Voronoi cells are selected based on a local version of the \emph{relative density} measure introduced in \cite{fortunato}, and an interpretable parameter $R$, called \emph{radius}, which tunes the scale (or resolution) at which the communities are detected, and can indirectly control the number of communities that are found. Since partitionings obtained by this method are controlled by a single scalar parameter, $R$, it is straightforward to select clusterings that maximize quality measures such as modularity. The method also allows for studying the hierarchical structure of the network, if it has one. The Voronoi community detection method is particularly advantageous when studying networks in which a natural connection length measure exists. Most other methods such as those based on the concept of modularity, require transforming these \emph{lengths} into a \emph{strength} measure first, which is often done in an ad-hoc manner. This feature of the algorithm is comparable to how centrality measures such as closeness or betweenness operate directly with connection lengths, in contrast to eigenvector centrality or PageRank which employ strengths instead. Another important feature of the algorithm is that using Voronoi partitioning the detected communties are contiguous,  a desirable property that does not hold for all methods, especially naive modularity maximization ones.

The algorithm was developed to be general and flexible. Depending on the data, and the interpretation of link weights, one can choose an appropriate transformation to convert connection weights into connection lengths, to be used by the Voronoi algorithm. 

We demonstrate applications of the algorithm on several real-world networks, namely: In the weighted and directed brain connectivity network of the macaque and the mouse, obtained by retrograde-tracing experiments \cite{cerebralcortex2012, plosbiology2016}, the weights are related to the probability of information transfer.  As a second example, the algorithm was used to find communities in a passenger air transportation network. In this example, the geographical distances between airports were also taken into account, as an illustration of the flexible way in which the Voronoi algorithm can employ distance data. Finally, we tested the algorithm on a random benchmark set that was generated using a weighted extension of the well-known Lancichinetti--Fortunato--Radicchi (LFR) algorithm \cite{PhysRevE.78.046110} (weights representing the strength of connections), and compared its performance to that of the Infomap algorithm. 

\section{Results}
\label{s:results}
\subsection{The algorithm\label{ss:algorithm}}
\textit{Graph Voronoi diagrams.} While most frequently Voronoi diagrams are defined in Euclidean or other metric spaces \cite{Voronoi_diagram_original}, they can also be defined on graphs \cite{VoronoiNJP, Erwig2000}. Let $G=(V, E)$ be a weighted directed graph consisting of set $V$ containing $N = |V|$ nodes/vertices and of set $E$ containing $M = |E|$ links/edges. Let us denote the weight of the link $i\rightarrow j$  by $w_{ij}$. In order to define Voronoi diagrams we must introduce the concept of length for each edge $i\rightarrow j$: $l_{ij}>0$. In directed graphs $l_{ij}$ and $l_{ji}$ are not necessarily equal. Depending on the graph the length of edges may depend on the weight $w_{ij}$ of links given in the dataset. As usual, the length of a path is obtained by summing up the lengths of edges along it. The distance $d(i,j)$ will denote the length of the shortest path going \textit{from} node $i$ \textit{to} node $j$. Here we focus on weighted directed graphs.

Let us choose a set of generator points (seeds) $S\equiv \{\gamma_1, \gamma_2, ..., \gamma_g\} \subset V$. We define the Voronoi diagram of $G$ with respect to $S$ as a partitioning of $V$ into disjoint subsets $V_1, V_2, ..., V_g \subset V$ called \emph{Voronoi cells}, where each cell is associated with a generator point, and they have the following two properties: (1) the Voronoi cells cover the original graph with no overlaps: $\cup_{i=1}^{g}V_i=V$ and $V_i\cap V_j=\emptyset$, $\forall i \ne j$; (2) nodes in a Voronoi cell are closest to the generator point of that particular cell: $d(n,\gamma_i)\leq d(n,\gamma_j)$, $\forall n \in V_i, i,j=1, 2, ..., g$. If, by a small chance, a node is at equal distance from two seeds we can choose its cell randomly between the two.

\textit{Distance measure}. 
While in unweighted networks one needs to consider only the topology of the network, here weights have a strong effect on community formation, posing also a greater challenge for detecting communities. We define the length $l_{ij}$ of link $i \rightarrow j$ based on two features of the network data:
(1)~Topological effects, i.e.\ the actual connectivity structure,  is incorporated by calculating the edge clustering coefficient $C_{n_i, n_j}$(ECC, see Methods) and taking $l_{ij}\sim 1/C_{n_i,n_j}$, as previously done in \cite{VoronoiNJP}.
(2)~The best way to take edge weights into account depends on the specific dataset and the interpretation of weights: do they represent a kind of connection strength, correlation measures, physical distances, bandwidth, flow measures, etc.?  For now let us write the relationship between the length $l_{ij}$ of a link and its weight $w_{ij}$ in a general form, as $l_{ij}\sim f(w_{ij})$.
The proposed length measure in this paper is a combination of the above mentioned two measures:
\begin{equation}
l_{ij}=\frac{f( w_{ij})}{C_{n_i, n_j} }.
\label{eq:wecc}
\end{equation}
The appropriate form of the $f(w_{ij})$ transformation depends on the dataset and needs to be determined by the user of the algorithm.  Some possibilities are the following.

1. In many cases, the weights already represent some sort of distance measure between the network nodes. For example, when the nodes are embedded in physical space, their geometric distances, $d_{ij}$ are already known. In such cases we may simply choose $f(d) = d$.
This approach was used in the case of air transportation networks (see Section \ref{ss:transpo}).

2. In some datasets, weights $w_{ij} \in (0,1]$ represent the probability of information flow along a given out-link of a node. A first approximation for this probability can often be obtained from correlation measures, normalized bandwidth of information transfer, or other measures. The probability for the information to travel along $n_i \rightarrow n_j \rightarrow n_k$ is proportional to $w_{ij} w_{jk}$. Then the logarithmic form of weights, $-\ln w_{ij} > 0$, is additive, and the shortest paths indicate routes where information flows with highest probability. This has been used in case of structural \cite{ErcseyRavasz2013} and functional brain networks \cite{Neuropharmacology}, state transition networks \cite{Entropy2021}, etc. In such situations we choose $f(w) = -\ln w$. In case of weights  representing correlation measures  the absolute value of the weights must be considered, because strong negative correlation also indicates strong information transfer, although such choice always depends on the data set itself and must be made on a case-by-case basis.
We illustrate this method on the inter-areal cortical brain network of the macaque monkey \cite{cerebralcortex2012} and the mouse \cite{plosbiology2016} (see Section \ref{ss:brain}).

3. In many cases, weights represent some type of connection strength, and it is desired that the algorithm should generally place nodes between which there is a high-weight link into the same community. Therefore, we need to choose a transformation that converts large weights into short lengths. A straightforward choice is $f(w) = 1/w$. In the current study, this methodology is used on a large set of benchmark networks with various degree distributions, densities, inter- and intra-community weight distributions (see Section \ref{ss:benchmark}).

\begin{figure}[h]
\includegraphics[width=1\textwidth]{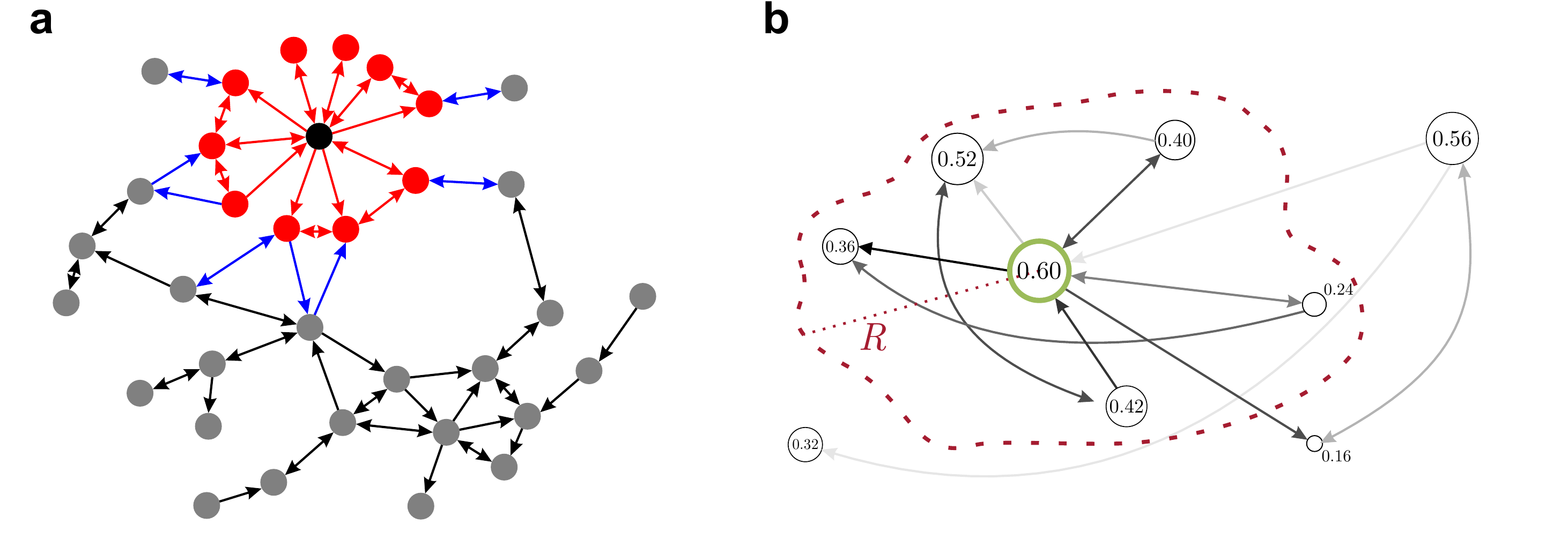}
\caption{Sketch of the algorithm. (a)~In order to calculate the weighted local node density $\rho_i$, we first count the links inside the first order neighborhood (red, $m=24$ in this example) and  the  links outgoing from and incoming into  the same neighborhood (blue, $k=12$ in this example). Links between two first order neighbors are also considered inside links (red). The unweighted local relative density would be $m/(m+k) = 2/3$, this needs to be multiplied by the weighted degree of node $i$. (b)~The size of nodes represents their weighted local relative density $\rho$, while the shade of links shows their weight $w$. Generator nodes (green border) are selected so that they have the highest weighted local relative density within a region of radius $R$ around them (red dashed line). No other generator points can exist within this region.}
\label{fig:network}
\end{figure}

\textit{Generator nodes.} The proper identification of the generator nodes is of key importance for obtaining Voronoi cells that are in good correlation with the community structure of the graph. The goal is to identify a generator point in each community such that the Voronoi cell induced by it would coincide with the community. It has been proposed in the past, both in the context of general clustering and community detection in networks, that clusters are typically concentrated around ``density peaks'' \cite{meng2021,Rodriguez2014}. Therefore, we generalize the \emph{local relative density} measure of \cite{VoronoiNJP} to a weighted context as follows:
\begin{equation}
\rho_i=s_i\frac{m}{m+k},
\label{eq:rho}
\end{equation}
where $s_i$ is its total (incoming and outgoing) weighted degree or \emph{strength}, $m$ is the number of edges within the first order neighborhood of node $i$, while $k$ is the number of edges entering into or exiting the same neighborhood. Links between two first order neighbor nodes are considered as inside edges, as shown in Figure \ref{fig:network}a. We refer to the quantity $\rho_i$  as the weighted local relative density of node $i$.

Node $i$ is chosen as generator point if it has the highest local relative density in a region within radius $R$: $\rho_i>\rho_j, \forall j\neq i , d(i,j)<R$. Algorithmically, to determine the generator points, first we select the node with the highest local relative density, then exclude all nodes from which it is no more than distance $R$ away. Then repeat the procedure on the remaining nodes until the entire network has been covered. This requires as many single-source shortest path calculations as the final number of generator points, determined by $R$. When using an optimal implementation of Dijkstra's algorithm \cite{dijkstra}, the procedure takes $O\bigl(g(|E| + |V| \log |V|)\bigr)$ operations, where $g$ denotes the number of generator points.

In networks that display a strong community structure, the inter-community node distances will be markedly smaller than the intra-community ones. This means that in order to detect communities accurately, it is important that each one should have exactly one generator point within it. However, the precise location of a generator point within its community does not influence the result significantly. We will demonstrate this on benchmark networks and the macaque brain network by randomizing the location of generator nodes (Section B5).

\textit{Node assignment to clusters.} After selecting the generator points $S \equiv \{ \gamma_1, \gamma_2, ..., \gamma_g \} \subset V$, we partition the nodes into clusters by constructing the graph's Voronoi diagram. Each node $i$ is assigned to the generator point $\gamma_k$ which is closest to it, i.e.\ $k$ is chosen so that $d(i,\gamma_k)$ is minimal (Figure \ref{fig:network}b).

\textit{Selecting the optimal clustering.} In order to choose an optimal partitioning of the graph, one may vary the $R$ scale parameter so as to maximize some quality measure. Here we chose the directed and weighted generalization of Newman's modularity for this purpose, as defined in \cite{newman_pnas}. The maximization in terms of the single $R$ parameter may be carried out with any derivative-free optimization technique. We chose Brent's method for its simplicity, bracketing the optimum between the graph diameter (largest possible $R$ value) and smallest edge length (smallest $R$ value).

\subsection{Structural brain networks}
\label{ss:brain}

\begin{figure}[h]
\includegraphics[height=1\textwidth]{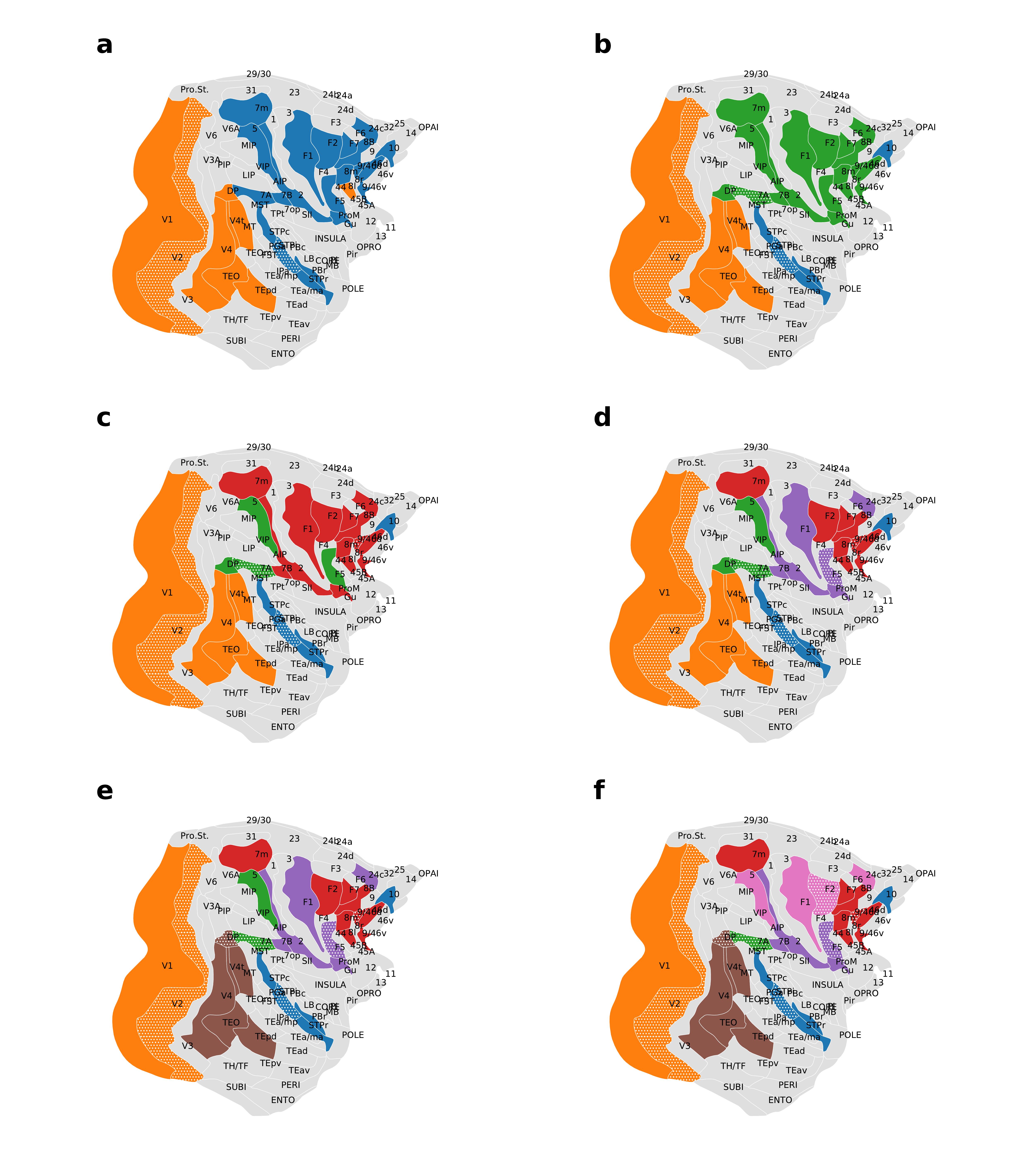}
\caption{Clustering of cortical brain areas in the macaque monkey based on their structural connectivity. Colors represent different clusters, while the generator nodes are marked with a dotted pattern. Connectivity data was unavailable for the light gray areas, thus these were excluded from the clustering. Panels (a)-(f) show partitionings into 2--7 clusters, as obtained with the Voronoi algorithm. The highest modularity is achieved with $7$ clusters.}
\label{fig:MonkeyFlatmap}
\end{figure}

\begin{figure}[h]
\includegraphics[height=1\textwidth]{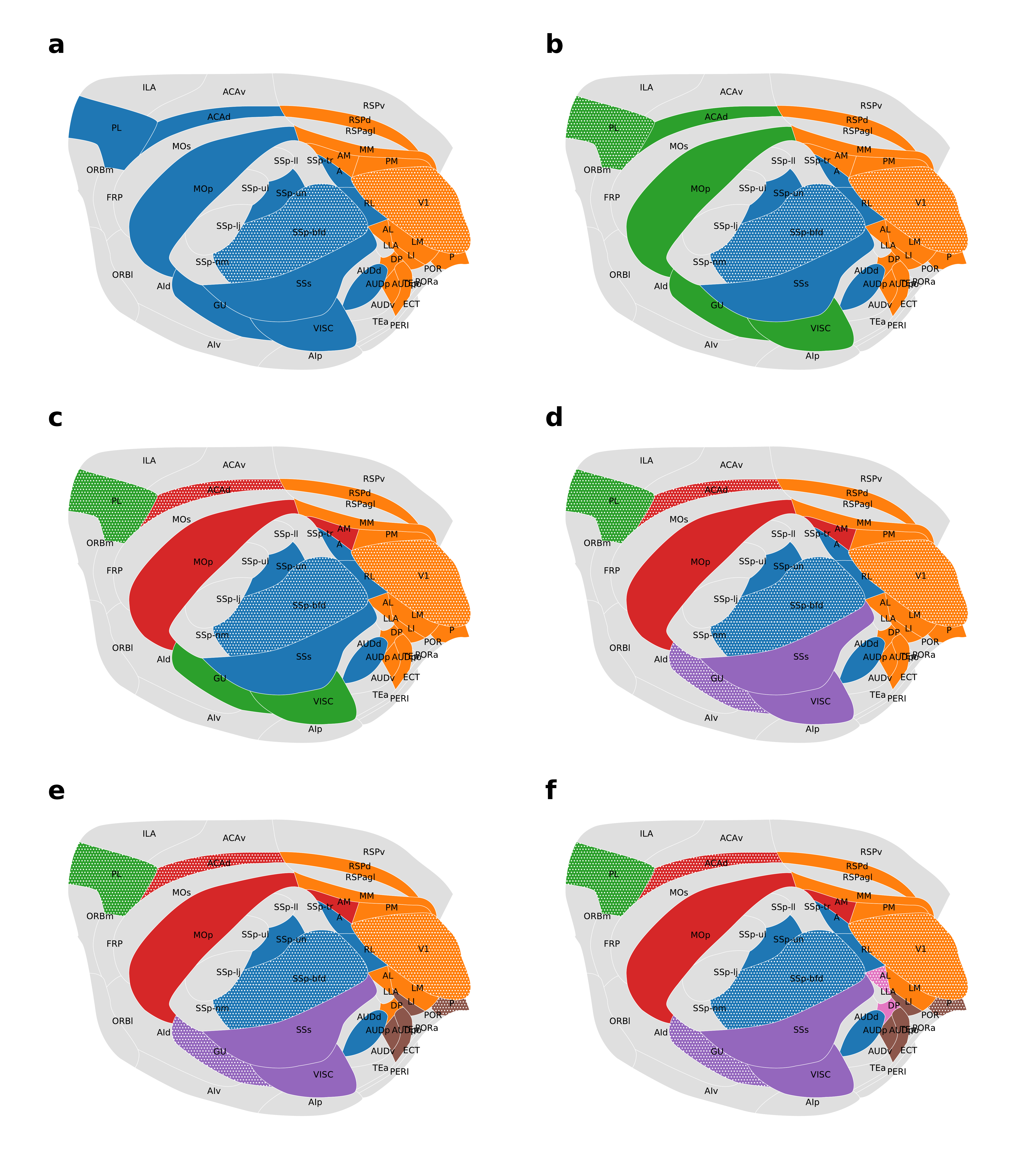}
\caption{Clustering of the brain network in case of the mouse. Colors represent different clusters, the patterns show the generator nodes of  clusters. The light gray functional areas are not included in the clustering due to the missing information about these areas. The Voronoi algorithm provided a fixed number of clusters, as follows: a) two; b) three; c) four; d) five; e) six; respectively f) seven clusters. In case of the mouse the optimal clustering is obtained with $2$ clusters.}
\label{fig:MouseFlatmap}
\end{figure}

First we demonstrate the clustering algorithm on anatomical brain connectivity data from the macaque \cite{cerebralcortex2012} and mouse brains \cite{plosbiology2016}. In this dataset, connection weights represent the relative fraction of connections incoming from each brain area. Since these are interpretable as probabilities for information transfer along given links, we choose the transformation $f(w) = -\ln w$. The macaque cortex  can be divided into 91 functional areas, out of which the connectivity of 29 areas was mapped in \cite{cerebralcortex2012}, meaning that only a subnetwork of $29$ nodes is completely known.  This network has a density of $66\%$ with a log-normal weight distribution and it was shown that weights carry important information, being dependent also on the physical distance according to the so-called exponential distance rule \cite{ErcseyRavasz2013}. The Voronoi clustering algorithm yielded  7 clusters in the optimal state (considering the highest Newman modularity) (Figure \ref{fig:MonkeyFlatmap}f) shown on the 2D brain flatmap. The mouse dataset contains the connectivity of 23 areas (out of $47$ in total), which form an almost complete graph ($96\%$ density). Thus, in this network, most information is carried by the connection weights. In this case the optimal number of clusters proposed by the algorithm is $2$ (Figure \ref{fig:MouseFlatmap}a). In Figures \ref{fig:MonkeyFlatmap} and \ref{fig:MouseFlatmap} functional areas with light gray color represent the areas which were not injected and are not included in the current clustering study. All other colors represent different clusters and the ones with patterns are the generator points for the particular cluster. To form a more complete picture several fixed cluster number states were also generated by the algorithm for both animals, ranging from $2$ clusters up to $7$  (Figure \ref{fig:MonkeyFlatmap}-\ref{fig:MouseFlatmap}a-f). We observe that by reducing the scale parameter $R$, large clusters break up into smaller ones, revealing the hierarchical structure of the brain. However, the data presented here is  incomplete, due to the missing information on the remaining functional areas. Including them would improve the accuracy of the clustering.

\subsection{Transportation networks}
\label{ss:transpo}

To further investigate our algorithm we tested it on air transportation networks provided by the Bureau of Transportation Statistics of the United States \cite{BTS}. The nodes of this network are airports and the links are airline routes. The database used in the current study is for the year $2018$ and contains data for passenger counts between any two airports in the United States. The available data is broken down into different airlines, months and days. The database also contains the great circle distance between the airports, which we denote by $d_{ij}$.

We calculated the edge weights $w_{ij}$ as the total yearly passenger count of the route divided by its geographic length, obtaining a value of dimension $\text{passenger} / \mathrm{km}$. This choice captures the intuition that if two routes of different lengths have the same passenger counts, then the shorter connection should be considered the ``stronger'' one. On short routes, one would generally expect smaller passenger counts since it is feasible to use other means of transport than air travel. The weights $w_{ij}$ obtained with this approach are only used for calculating the local relative density of nodes, and, at the end of the clustering process, determining the weighted modularity of the partitioned network at different values of radius $R$. To compensate for the incompleteness of the database, only the strongly connected giant component of the constructed network is used in this study, consisting of $1106$ airports. 

The length of links used in the Voronoi partitioning were chosen solely based on the $d_{ij}$ geographical distances, excluding the passenger count data: $l_{ij}=f(d_{ij})/C_{n_i,n_j}=d_{ij}/C_{n_i,n_j}$.

\begin{figure}[h]
\includegraphics[width=1\textwidth]{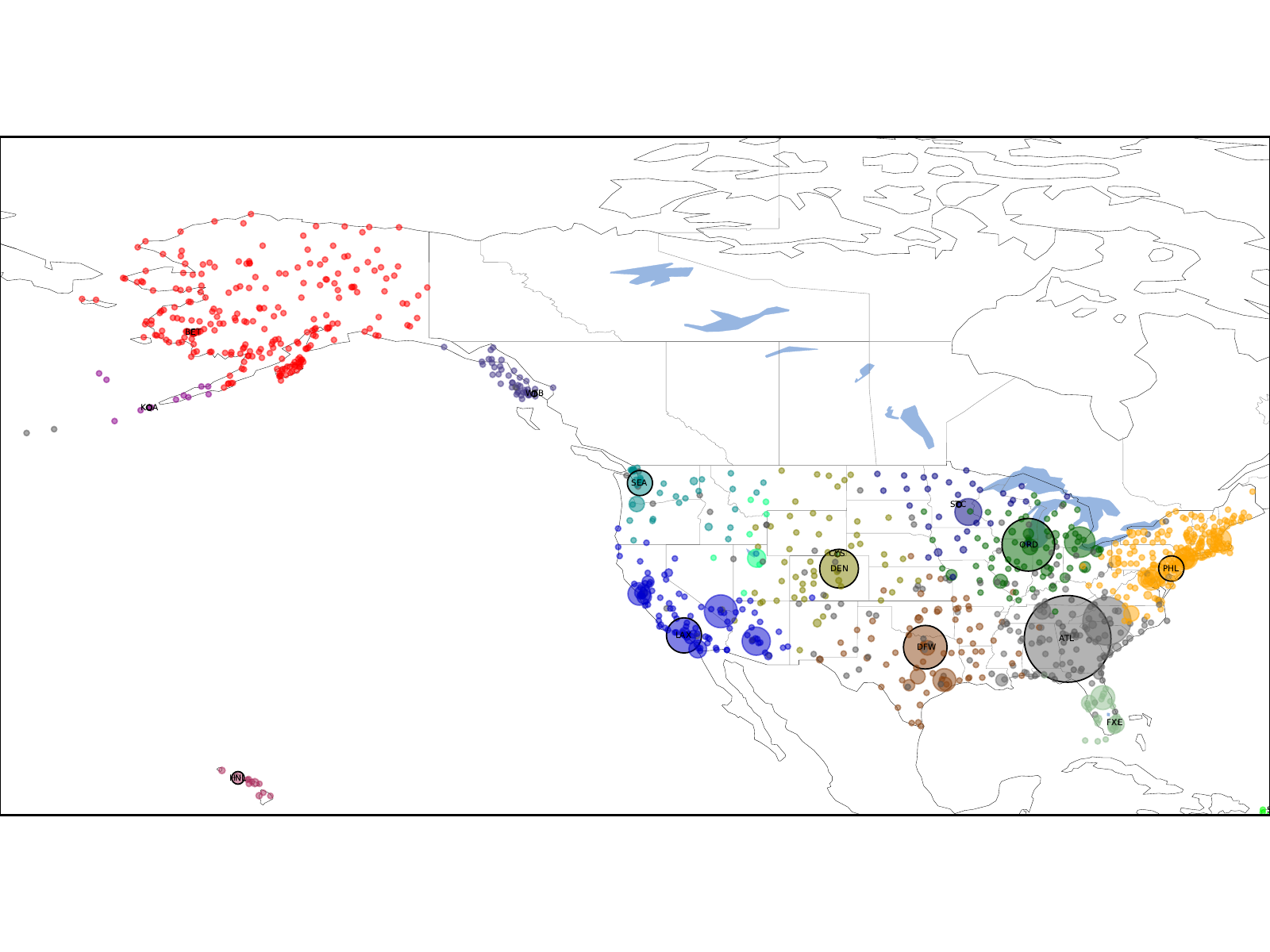}
\caption{Clustering of the air passenger transportation network of the United States using the Voronoi algorithm, yielding $116$ clusters, comprising $14$ large clusters and $102$ small clusters. The disks represent the locations of airports on the map. Their size is proportional to their local relative density, and their color indicates the cluster they belong to. Generator nodes are shown with a black border. Only generator point airports are annotated with the IATA airport code.}
\label{fig:dperecc}
\end{figure}

The Voronoi algorithm yields an optimal clustering consisting of the $116$ communities shown in Figure \ref{fig:dperecc}. Of these, only $14$ are large enough to be clearly visible in the figure, but there are also $102$ tiny clusters, often consisting of a single node. These tiny clusters appear because   the database is often incomplete (number of links, passenger counts etc.), especially in case of small airports, which often become their own isolated cluster. This drawback can be corrected by using a more complete database, which was not available at the time of this study. However, examining only the $14$ large clusters one can conclude that the algorithm separates extremely well the airports of different major US geographical regions: the East Coast (orange), Midwest (light blue), West Coast (brown), Alaska (cyan), Southeast (light orange) and Central around Atlanta (blue) regions. Often, the selected generator points for these clusters are the major airport hubs of that particular region, e.g.\ Hartsfield-Jackson Atlanta International Airport (ATL), Chicago O'Hare International Airport (ORD), etc.

\begin{figure}[htb]
\includegraphics[width=1\textwidth]{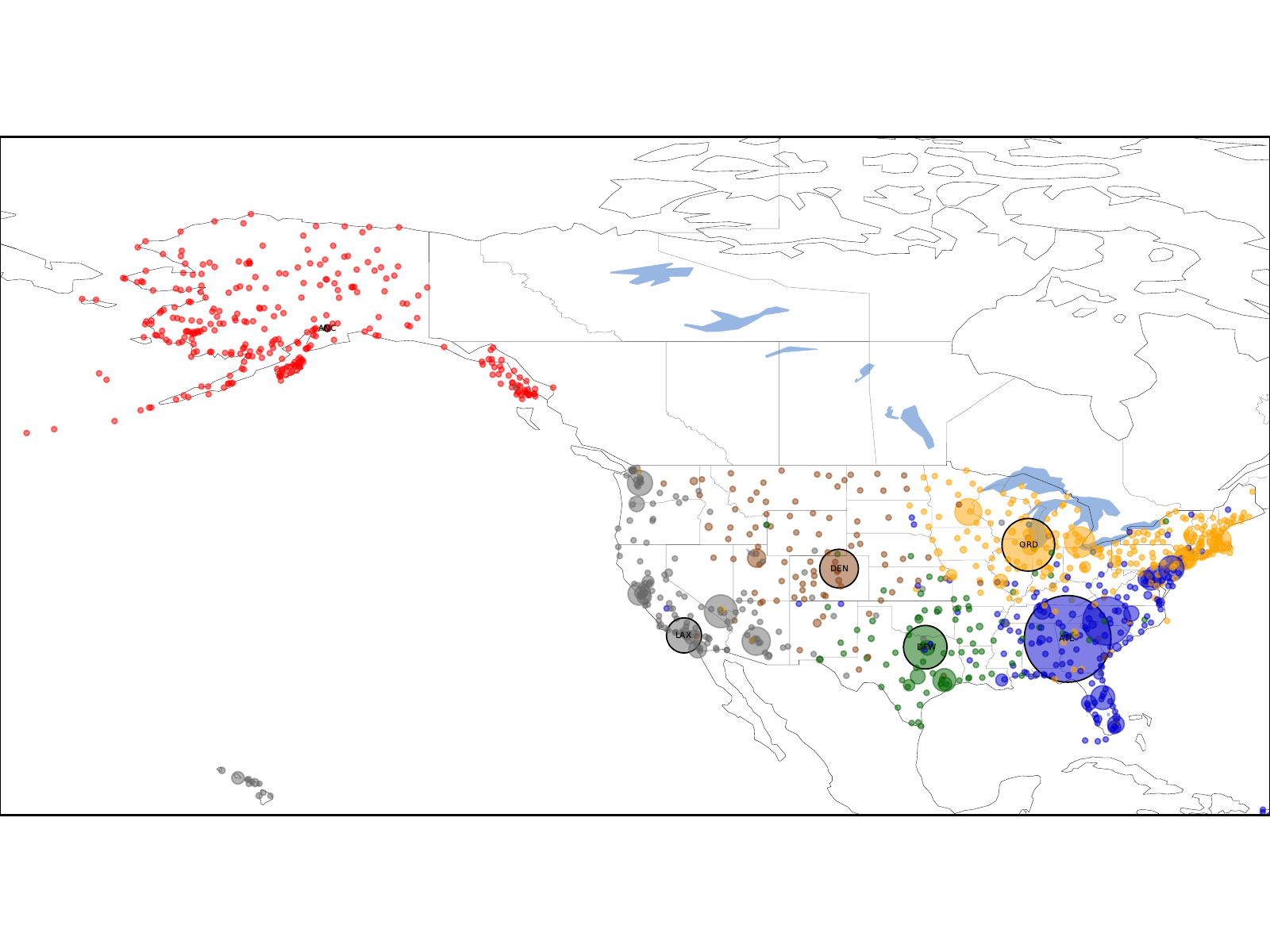}
\caption{Clustering of the air passenger transportation network of the United States using the Voronoi algorithm using the Top 5 busiest airports (ATL, LAX, ORD, DFW, DEN) and ANC (Anchorage, Alaska) as generator points. The disks represent the locations of airports on the map. Their size is proportional to their local relative density, and their color indicated the cluster they belong to. Generator nodes are shown with a black border. Only generator point airports are annotated with the IATA airport code.}
\label{fig:top5ak}
\end{figure}

Another key feature of our algorithm is its ability to handle predefined generator points. To show this, in Supplementary Figures 1 and 2 the top 5, respectively the top 10 busiest US airports by total passenger traffic were pre-set as generator points. Also, as previously shown in Figure \ref{fig:dperecc}, Alaska has its trend to form its own cluster, so two additional cases were investigated, where Ted Stevens Anchorage International Airport (ANC) was included as an additional generator point besides the top $5$ and top $10$ busiest airports (Figure \ref{fig:top5ak} and Supplementary Figure 3). The $10$ busiest airports based on total passenger count in $2018$ in the United States are as follows: Hartsfield-Jackson Atlanta International Airport (ATL), Los Angeles International Airport (LAX), O'Hare International Airport (ORD), Dallas/Forth Worth International Airport (DFW), Denver International Airport (DEN), John F. Kennedy International Airport (JFK), San Francisco International Airport (SFO), Seattle-Tacoma International Airport (SEA), McCarren International Airport (LAS) and Orlando International Airport (MCO). This list is based on annual  passenger traffic figures published for 2018 by each airport authority. From all four figures (Fig. \ref{fig:top5ak} and Supplementary Figures $1-3$) one can draw two major conclusions: (1)~the optimal clustering found by the Voronoi algorithm is in good agreement with the catchment area of the biggest airports; (2)~the introduced local relative density measure is proportional with the real life data of actual annual traffic of each airport.

\subsection{Results on randomly generated benchmark networks} \label{ss:benchmark}

To demonstrate the power of the algorithm, extensive tests were performed on random directed benchmark networks generated using the popular LFR algorithm \cite{Fortunato_Lancichinetti}, which outputs both a network and an associated \emph{ground truth community structure}. In order to be able to investigate the effect of different link weight distributions on the Voronoi algorithm's performance, weights were sampled in a second, separate step, from either a normal distribution or a power distribution. Different distribution parameters were used for inter- and intra-community links, ensuring that links contained within communities would have an average weight at least as high as those connecting different communities. See Methods \ref{ss:bencharkgenerator} for the details of the generation procedure.

\begin{figure}[h]
\includegraphics[width=0.7\textwidth]{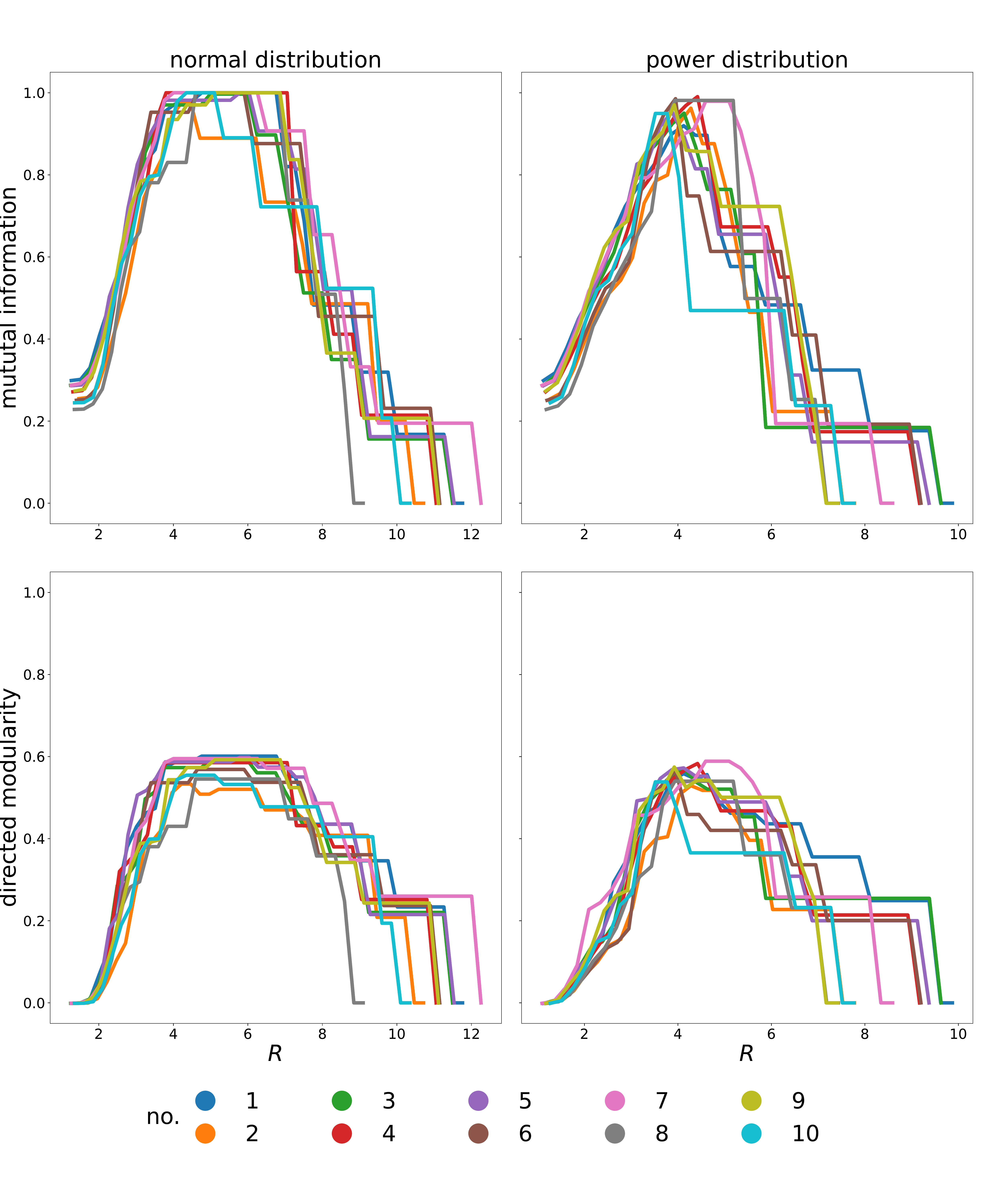}
\caption{
Mutual information (MI) between the clustering obtained by the Voronoi algorithm and the ground-truth clustering encoded in the benchmark networks as function of radius $R$ (upper row) and directed Newman modularity as a function of radius $R$ (lower row) in case of normal (left column) and power (right column) weight distributions of the same $10$ (see legend) networks with $N=1000$,  $k = 100$ and mixing parameter $\mu=0.3$. The base network is the same for both distribution types, only the weights differ according to the distribution parameters chosen to have the same distribution mean in both cases of normal and power distributions: $m_{intra}=0.58, m_{inter}=0.42$ for normal, while $\alpha_{intra}=0.7, \alpha_{inter}=0.3$ for power distributions.
} 
\label{fig:mutualinfo}
\end{figure}

Since weight values represent the strength of each link in this case, we used the $f(w_{ij})=1/w_{ij}$ transformation when calculating the length of links.
Figure \ref{fig:mutualinfo} shows the mutual information between the ground truth community structure and the detected one, as well as the directed modularity values, as a function of the $R$ parameter in $10$ networks with $N=1000$ nodes, $\overline{k}=100$ average degree ($k_{max}=300$)  and mixing parameter $\mu=0.3$. Most curves show a plateau around their maximum, indicating that the optimal clustering can be achieved for a range of $R$ values.

\begin{figure}[h]
\includegraphics[width=0.7\textwidth]{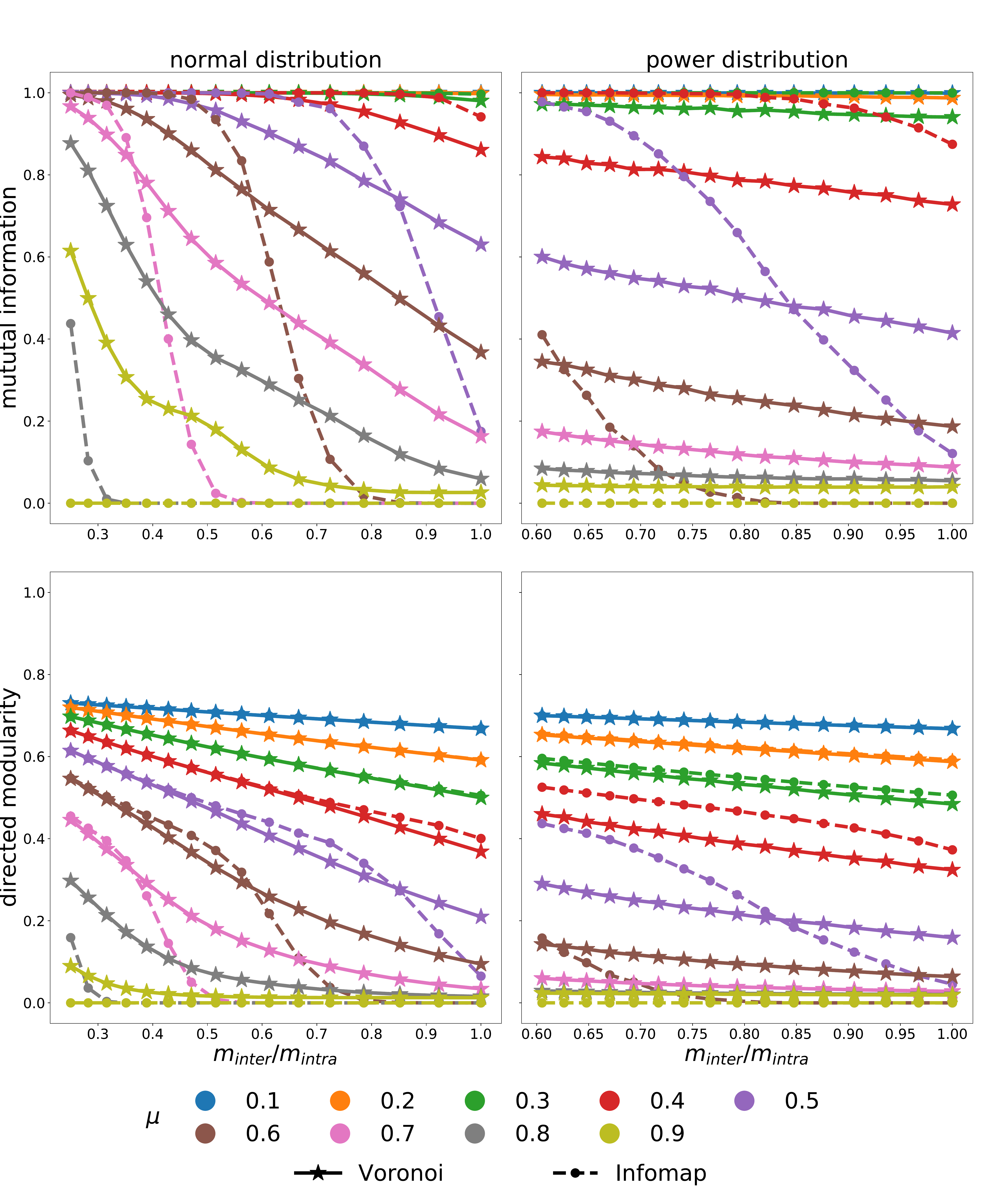}
\caption{Performance comparison of the Voronoi algorithm (solid line, stars) and Infomap (dashed line, dots) using a total of $288\,000$ benchmark networks. The mutual information (MI) between the detected communities and the ground truth, as well as the modularity of the detected communities is shown as a function of the weight ratio between inter- and intra-community links. Curves of different colors correspond to LFR benchmark networks generated with different mixing parameters ($\mu=0.1, \dots, 0.9$). All benchmark networks have $N=1000$ nodes and mean degree $\bar k = 100$. The parameters of the link weight distributions were chosen for both, normal and power distributions, as follows: $(m_\text{inter}, m_\text{intra}) = (0.20, 0.80), (0.22, 0.78) \dots, (0.50, 0.50)$. Each data point is averaged over $1000$ LFR networks. Infomap tends to exhibit an almost binary behaviour in its accuracy: it either detects communities perfectly (indicated by an MI value of 1) or fails to detect them at all. In contrast, the Voronoi algorithm shows a more gradual transition in its accuracy, and will find a reasonable community structure even in difficult benchmark instances.
}
\label{fig:bench}
\end{figure}

Figure~\ref{fig:bench} compares the performance of the Voronoi algorithm with that of Infomap, which is one of the most popular community detection methods used with weighted and directed networks today. Each curve shows comparisons done for benchmark networks created with a given mixing parameter. The higher the mixing parameter, the less separated, and thus the harder to detect the communities are. The inter- and intra-community weight distributions are varied from more distinct to more similar in the figure panels from left to right, with the horizontal axis showing the ratio of the average inter- and intra-community weights. The smaller this ratio, the easier it is to detect communities accurately. We observe that while both community detection methods perform well on easy benchmark instances, they differ in how their accuracy decreases as the problem gets more difficult. Infomap exhibits a binary behaviour: it either recovers the ground truth community structure nearly perfectly (indicated by a mutual information value of 1, see top panels), or it fails to find any communities at all (a mutual information of 0). Closer examination reveals that when this happens, Infomap either returns a single community, or places each vertex in its separate size-1 community. In contrast, the accuracy of the Voronoi algorithm decreases gradually, and it still manages to recover a reasonably accurate community structure on difficult problem instances. The mutual information measure, plotted in the top panels, only indicates how good the agreement is between the ground truth and the detected communities, but it does not tell us how well-defined the community structure is. For this reason, we also plot the directed modularity of the detected communities in the bottom panels, and demonstrate that a reasonably strong community structure is indeed present in many of the benchmarks instances where Informap failed to find any.

A speed performance comparison was conducted between the Voronoi and Infomap algorithms on an Apple MacBook Pro (16 inch, 2019) laptop using $1000$ randomly generated benchmark networks with $N=1000$ nodes, average degree $\overline{k}=100$, mixing parameter $\mu=0.3$, with power law weight distribution having $\alpha_{intra}=0.6$ for intra-community, respectively $\alpha_{inter}=0.4$ for inter-community edge weights. For the Voronoi algorithm it takes on average $\overline{t}_\text{Voronoi}=0.32\;\mathrm{s}$ to find an optimal clustering of one benchmark clustering problem. The Infomap algorithm averages $\overline{t}_\text{Infomap}=1\;\mathrm{s}$ per problem.

\subsection{Robustness of the generator point selection}

To demonstrate the efficiency and robustness of the generator point selection, we investigated Voronoi partitions obtained after a random rearrangement of the generator points. We consider two types of rearrangements. In the first approach, an initial Voronoi clustering is constructed using generator points selected in the usual manner, based on high local relative density, and finding the optimal $R$ parameter value. Then each generator point is replaced by a node selected randomly from within its cluster. We refer to this as \emph{intra-community} randomization. In the second approach, the same number of generator points are sampled uniformly at random (without replacement) from the set of all nodes. We call this \emph{uniform} randomization. Finally, a new Voronoi clustering is obtained from the rearranged generator points.

\begin{figure}[htb]
\includegraphics[width=.66\textwidth]{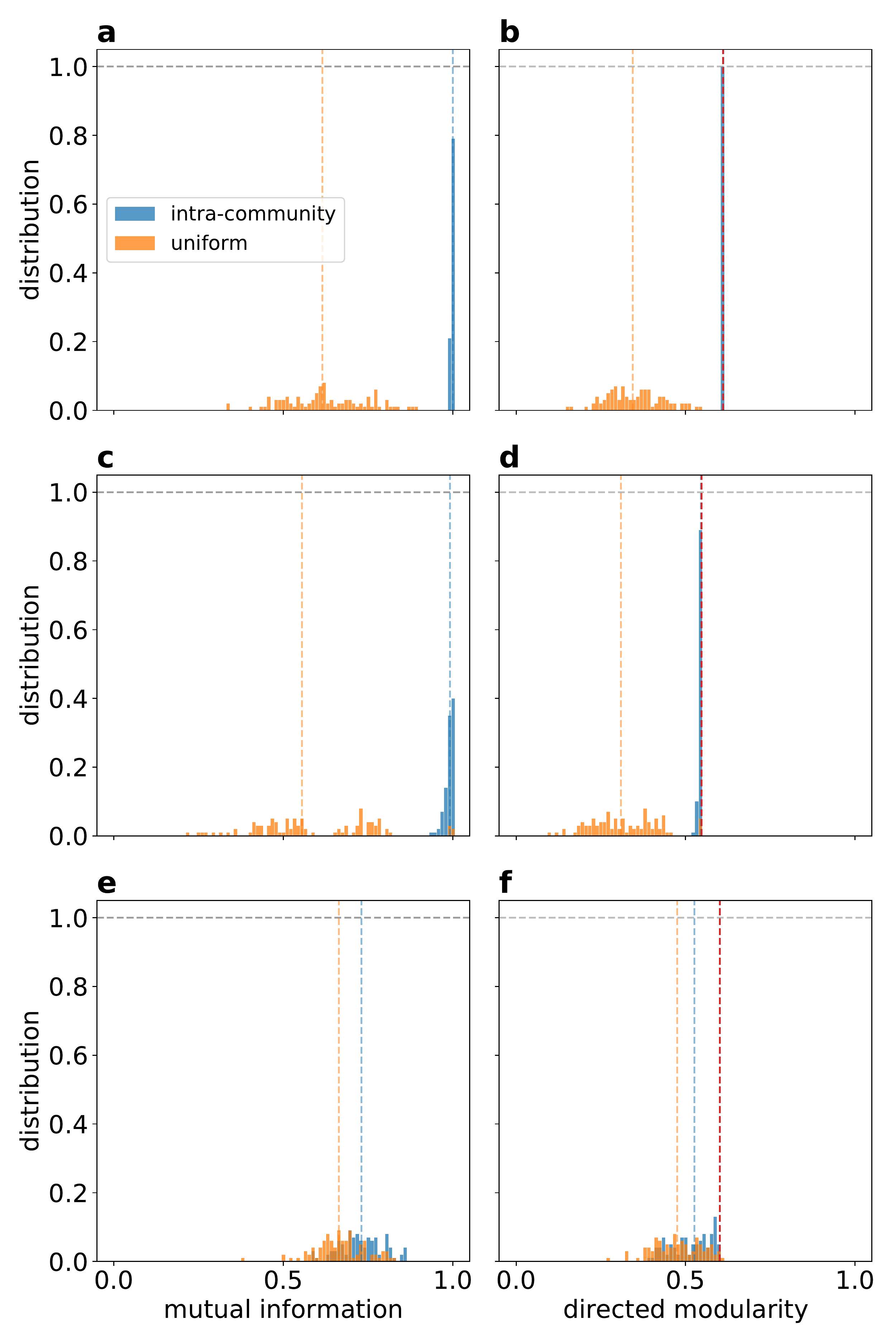}
\caption{Histogram of mutual information (\textbf{a, c, e}) and directed modularity (\textbf{b, d, f}) values for 100 randomized generator point set in case of: (\textbf{a, b}) a benchmark network with higher ground truth directed modularity value ($Q=0.610638$); (\textbf{c, d}) a benchmark network with lower ground truth directed modularity value ($Q=0.547427$) and (\textbf{e, f}) the cortical functional network of the macaque monkey. Both random benchmark networks have $N=1000$ nodes, an average degree $\overline{k}=100$, mixing parameter $\mu=0.3$ and the link weights follow a power distribution with $\alpha_\text{intra}=0.7$ and $\alpha_\text{inter}=0.3$ for intra- and inter-community links, respectively.
Blue bars represent the intra-community randomization of the generator points, while the orange ones the uniform randomization. The vertical blue and orange dashed lines show the median of the distribution, while the vertical red dashed line on the directed modularity plots represents the directed modularity value of the ground truth clustering in case of the benchmark networks and the optimal state modularity obtained by the Voronoi algorithm in case of the cortical network of the macaque.}
\label{fig:misplacedAndRandom}
\end{figure}

For this experiment, we used two LFR benchmark networks as well as the macaque brain connectivity network. The two benchmark networks were both generated using the same LFR parameters ($N=1000$ nodes, average degree $\overline{k}=100$ and mixing parameter $\mu=0.3$) and the same weight distributions (power law distributions with exponents $\alpha_\text{intra}=0.7$ for intra-community and $\alpha_\text{inter}=0.3$ for inter-community links). The two benchmark networks were specifically selected from a large randomly generated set so that their ground truth community structures would have very different modularity values ($Q=0.610638$ in Fig.~\ref{fig:misplacedAndRandom}a,b and $Q=0.54742$ in Fig.~\ref{fig:misplacedAndRandom}c,d).

Each type of randomization was repeated $100$ times on these three networks, and the obtained clusterings were evaluated using the modularity and mutual information metrics. In case of the benchmark networks, the ground truth community structure was used as the baseline of the evaluation. Since no ground truth is available for the macaque brain network, the optimal Voronoi clustering was used in its place. The obtained results are shown in Figure \ref{fig:misplacedAndRandom}, where light blue colors represent the results obtained after intra-community randomization, while the orange shows the results for uniform randomization. Red dashed lines indicate the baseline modularity.

With both the high and low modularity benchmark networks, the Voronoi algorithm, in its original form, was able to recover the ground truth clustering exactly, as indicated by a mutual information value of 1. Intra-community randomization had very little effect on the result (blue). This demonstrates that when precisely one generator point is placed into each community, the Voronoi approach is accurate in recovering the true community structure, regardless of the specific location of the generators. The local relative density-based generator selection method is very effective in achieving this one generator per community arrangement. In comparison, a random placement of generators leads to a significantly worse result (orange).

In case of the macaque brain network, the obtained clustering is more sensitive to intra-community randomization. However, it must be noted that this 29-node network is much smaller than the 1000-node benchmark graphs, more dense, so topologically less modular (its modular nature is hidden more in the link weights)  therefore large fluctuations are to be expected. Even so, intra-community randomization produces clusterings with clearly higher modularity than uniform randomization, which supports the original choice of generator points, based on local relative density. Our original method shown as the baseline (red dashed line) provides almost the maximal modularity achieved on this network.

\section{Discussion}

In this paper we presented a computationally efficient method for detecting community structure in directed and weighted networks. An advantage of our algorithm is that it can make direct use of edge weights which represent additive \emph{distances}, i.e.\ data in which larger weights indicate weaker connections. Most other community detection methods, including all of those that maximize modularity explicitly, interpret weights as \emph{connection strengths}, and thus require transforming distance data before they can be used. This transformation is often done in an ad-hoc manner. In contrast to such methods, Voronoi community detection provides a principled way to work with distance-weighted networks. This feature is similar to how some network metrics, such as betweenness centrality, employ distances, while others, such as PageRank, use connection strengths.

While Voronoi community detection is a natural fit for distance-weighted data, it can also work with weights of different types after applying a suitable weight transformation. The most appropriate transformation depends on the specific use case. We discuss three possible approaches to this, which covers many different types of networks.

Regarding directions, we have chosen to use the outgoing  paths (defining the region with radius $R$ with shortest paths outgoing from a node) when choosing generator nodes, and also when deciding the community of each node they belong to (calculating the shortest path starting from the node that needs to be assigned to a cluster to the different generator nodes). This choice is motivated by networks that carry some kind of flow, for example information flow in brain networks. One may intuitively think of nodes with strong incoming connections as points of attraction, and the Voronoi cell of a generator point as its sphere of attraction. However, the same method may also be used with incoming paths, or even compare the community structure obtained in the two different cases.

As we have shown in case of benchmark networks, there is typically a range of $R$ radius values which yield the optimal clustering, which helps the time performance of the algorithm when searching for the optimal $R$ value. The most computationally expensive part of the algorithm is calculating shortest path lengths in the graph. This can be realized with Dijkstra's algorithm \cite{dijkstra} which has complexity $\mathcal{O}(|E|+|V|\log |V|)$ if optimal implementation is used. In case of graphs with high density, the simpler-to-implement Floyd--Warshall algorithm ($\mathcal{O}(|V|^3)$) \cite{FloydSP, WarshallSP} also provides good performance.

We contributed an implementation of the Voronoi community detection algorithm (\texttt{igraph\_community\_voronoi()}) and its building blocks (\texttt{igraph\_ecc()} and \texttt{igraph\_voronoi()}) to the \emph{igraph} open-source network analysis software, making the method easily available to practitioners of network analysis. 

\section{Methods}
\label{s:methods}

\subsection{Edge Clustering Coefficient}
\label{ss:ecc}
In the weight to length transformation formulae $C_{n_i, n_j}$ denotes the value of the edge clustering coefficient (ECC) of edge $(n_i, n_j)$, defined in \cite{ECC} as
\begin{equation}
C_{n_i, n_j}=\frac{z(n_i, n_j) + 1}{\min[k(n_i) - 1, k(n_j) - 1]},
\label{eq:ecc}
\end{equation}
where $z(n_i, n_j)$ denotes the number of common neighbors of nodes $n_i$ and $n_j$, i.e.\ the number of triangles in which the $(n_i, n_j)$ edge participates, while $k(n_i)$ and $k(n_j)$ are the degrees of $n_i$ and $n_j$. Edge weights and edge directions are not taken into consideration when computing this measure.

\subsection{Modularity}
\label{ss:modularity}
In this study we used the weighted and directed version of Newman's modularity \cite{Leicht2008}, defined as
\begin{equation}
Q=\frac{1}{m}\sum_{i,j\in V}\left[w_{ij}-\frac{s_i^\text{out}\cdot s_j^\text{in}}{m}\right]\cdot\delta_{c_i, c_j},
\label{eq:modularity}
\end{equation}
where $m = \sum_{i,j} w_{ij}$ denotes the total sum of edge weights in the graph, $s_i^\text{out}$ and $s_j^\text{in}$ represent the weighted out- and in-degree of nodes $i$ and $j$, while $\delta_{c_i, c_j}$ is the Kronecker symbol, which equals to 1 if nodes $i$ and $j$ are in the same cluster, $c_i=c_j$, and $0$ otherwise. 
Many studies argue \cite{newman_pnas, Newman_Networks, newman_girvan, girvan_newman} that optimizing modularity is a good way of detecting community structure, however, it is not universal as shown by \cite{louvain}.

\subsection{Generating benchmark networks}
\label{ss:bencharkgenerator}
Benchmark networks used in the current study were created based on the popular LFR algorithm introduced in \cite{PhysRevE.78.046110,LFRcode}, which generates a random network with a predefined community structure. Here we used the original implementation for directed graphs, written by the algorithm's authors \cite{LFRcode}.  The LFR benchmark generator requires specifying a so-called mixing parameter $\mu \in (0, 1)$, which controls how well the generated communities are separated from each other, in other words, how easy it is to distinguish between the clusters. Specifically, a fraction $1-\mu$ of links is contained within communities. As the value of the mixing parameter is increased, the clusters are less separated, and it becomes harder to detect them accurately. We generated benchmarks sets corresponding to the full range of mixing parameter values $\mu = 0.1, 0.2, \dots, 0.9$. The other parameters of the LFR benchmark model were fixed as $N=1000$ nodes, average degree $\bar k = 100$, maximum in-degree $k_\text{max} = 300$.

Although the benchmark network generator software \cite{LFRcode} is capable of producing link weights, we only used it to create binary (unweighted) networks and sampled the edge weights separately. This way we were able to study how different weight distributions influence the performance of the Voronoi algorithm. Two different types of edge weight distributions were considered: (1)~Truncated normal distribution with probability density $p(w) \sim e^{-\frac{1}{2} \left(\frac{w-m}{\sigma}\right)^2}$, $w \in (0, \infty)$; and (2)~power distribution with probability density $p(w) \sim \alpha\, w^{\alpha - 1}$, $w \in (0, 1]$ and $0 < \alpha \le 1$. Since the power distribution assigns very high probability to tiny values which may become rounded to zero during floating point computations, its support was truncated to $w \ge w_\text{min} = 0.01$. The expected value of this truncated power distribution is $m = \frac{\alpha}{1+\alpha} \frac{w_\text{min}^{\alpha+1} - 1}{w_\text{min}^\alpha}$.

The steps for generating the random weighted and directed benchmark networks are the following: (1)~generate a binary directed network with a specific mixing parameter $\mu$; (2)~choose the type of edge weight distribution (normal or power); (3)~sample the edge weights, choosing different distribution parameters for inter- and intra-community edges, so that the average weight of links within communities would be no smaller than that of links connecting distinct communities, $m_\text{intra} > m_\text{inter}$. Using these steps we ensure that the weights will not overwrite the ground truth community structure defined in the first step.

\subsection{Mutual information}
\label{s}
The \emph{mutual information} (MI) measure can be used to characterize the similarity between two different partitionings of the same network, denoted $A$ and $B$. It is defined as $MI(A, B) = H(A) + H(B) - H(A, B)$, where $H(A)$ and $H(B)$ are the entropies of the individual partitonings, while $H(A, B)$ is their joint entropy.  Intuitively, it characterizes the ``overlap'' between the two partitionings. For a given $A$, it reaches its maximum when $B$ becomes identical to $A$, and a value of zero when the membership of a node in a particular cluster within $A$ conveys no information about its position within $B$. Note that the MI is invariant to a relabeling of the clusters.

In order to calculate the MI let us assume that there are $r$ clusters within $A$, and $c$ clusters within $B$, respectively. Based on $A$ and $B$ one can construct a contingency table: element $n_{ij}$ of this table gives the number of nodes that belong to cluster $i$ of $A$  and cluster $j$ of $B$, as shown in Table~\ref{table:contingency}.
\begin{center}
\begin{table}[h!]
    \centering
     \begin{tabular}{|c|ccccc|}
    \hline
        & $b_1$ & $\ldots$ & $b_j$ & $\ldots$ & $b_c$\\ \hline
  $a_1$ & $n_{1,1}$ & $\ldots$ & . & $\ldots$ & $n_{1,c}$\\
  $\vdots$ & $\vdots$ &  & $\vdots$ & & $\vdots$\\
  $a_i$ & . &  & $n_{i,j}$ &  & .\\
  $\vdots$ & $\vdots$ &  & $\vdots$ & & $\vdots$\\
  $a_r$ & $n_{r,1}$ & $\ldots$ & . & $\ldots$ & $n_{r,c}$\\
    \hline
    \end{tabular}
    \caption{$r\times c$ contingency table for the two different cluster assignments ($A$ and $B$) of the same network, where $a_i = \sum_j n_{ij}$, while $b_j = \sum_i n_{ij}$.}
    \label{table:contingency}
\end{table}
\end{center}
Then the entropies of the two partitionings are calculated as
\begin{equation}
 H(A) = -\sum_{i=1}^{r} \left(\frac{a_i}{N} \, \ln \frac{a_i}{N}\right)
\end{equation}
\begin{equation}
 H(B) = -\sum_{j=1}^{c} \left(\frac{b_j}{N} \, \ln \frac{b_j}{N}\right),
\end{equation}
and the the mutual information is
\begin{equation}
\label{eq:mi}
MI = \sum_{i=1}^{r}\sum_{j=1}^{c} \left(\frac{n_{ij}}{N} \, \ln \frac{n_{ij} \, N}{a_i \, b_j}\right).
\end{equation}
Here were normalize the mutual information as $NMI = \frac{MI}{\max\{H(A), H(B)\}}$ in order to obtain a value within the interval $[0, 1]$. A values of 1 is achieved precisely when $A$ and $B$ coincide.

\subsection{Computing a Voronoi partitioning}

In order to compute the Voronoi partitioning of a graph, one must find the closest generator point to each vertex. From an algorithmic perspective, the simplest way to do so is to pre-compute the distances between all pairs of vertices using the Floyd--Warshall algorithm, which is simple to implement, but takes time proportional to $|V|^3$, where $|V|$ represents the size of the graph's vertex set. However, when the number of generator nodes is small, the calculation can be performed more efficiently using a single-source shortest path algorithm, applied once to each generator point. We used Dijkstra's algorithm for this, implemented using a binary heap, which takes time proportional to $|S| \, |E| \log |V| + |V|$, where $|S|$ and $|E|$ are the sizes of the generator set and the graph's edge set, respectively. An additional optimization is possible by limiting the shortest path search from each generator up to those vertices which are not closer to any previously processed generator point.  This way, the shortest path search will explore successively smaller and smaller regions of the graph with each new generator point, leading to a sub-linear complexity in the number $|S|$ of generators. On average, we expect to explore a fraction $1/k$ of the graph when processing the $k$th generator, therefore the time to compute the entire Voronoi partitioning will be proportional to $\sum_{k=1}^{|S|} 1/k \sim \log |S|$, yielding a final computational complexity of $O(\log |S| \, |E| \log |V| + |V|)$. The logarithmic scaling in the number of generators allows us to efficiently handle a large number of generator points, which is necessary when maximizing quality metrics of the partitioning in terms of the radius parameter $R$.

\begin{acknowledgments}
\section{Acknowledgements}
We are grateful to the research group of Henry Kennedy, especially to Răzvan Gămănuț and Loïc Magrou, for providing the mouse and macaque brain flatmaps to illustrate the results.
This work was supported by the Human Brain Partnering Projects COFUND-FLAGERA II-CORTICITY and COFUND-FLAGERA-ModelDXConsciousness (M.~E.-R.), a grant by the European Union Horizon 2020 Research and Innovative Program - grant agreement no. 952096  (NEUROTWIN), 
grants of the Romanian Ministry of Research, Innovation and Digitization, CNCS/CCCDI -
UEFISCDI, project number ERANET-NEURON-2-UnscrAMBLY,  ERANET-FLAG-ERA-ModelDXConsciousness, and PN-III-P4-PCE-2021-0408 within PNCDI III (M.E.-R.),  and by a Young Researcher's Grant of the StarUBB Institute, provided by the Babeș-Bolyai University (GTC 31377/2020) (B.~M.). Sz.~H. is grateful to the Wissenschaftskolleg zu Berlin for the time provided to work on this project.
\end{acknowledgments}

\section{Author contributions}
B.~M., Sz.~H. and M.~E.-R. designed the methods and algorithm. B.~M., I.~B.~M. and Sz.~H. implemented and tested the algorithm. B.~M and I.~B.~M performed data analysis. B.~M, Sz.~H. and M.~E.-R. wrote the manuscript, all co-authors reviewed the manuscript.

\section{Additional information}
\textbf{Competing interests} The authors declare no competing interests.


%

\end{document}